# Optical vortex trajectory of the edge-diffracted single-charged Laguerre-Gaussian beam

**Aleksey V. Chernykh and Nikolay V. Petrov**

ITMO University, St. Petersburg 197101, Russia

E-mail: chernikh.a@gmail.com

**Abstract**

The paper deals with the edge diffraction of the single charged Laguerre-Gaussian beam outside the waist. Based on the Kirchhoff-Fresnel integral, the behavior of the optical vortex (OV) migration during sequential beam blocking by the straight edge of the screen is performed analytically. Universal combination of the diffracting-beam parameters determining the shape for the OV spiral trajectories is revealed. Parametric equations describing the OV dislocation dependent on the beam screening degree are derived. In addition, the spiral trajectory, which describes the OV positions in the case of the screen-edge translatory motion across the incident beam under conditions of the weak diffraction perturbation, is obtained. As a result, the equation for a wider region of diffraction perturbation is selected as a best matching function for a certain variation interval of the beam parameters. The work has prepared the basis for solving the inverse diffraction problem: determining parameters of the beam diffractive interaction with obstacles from the migration trajectory of the optical vortex.

**Keywords**: optical vortex, diffraction, Laguerre-Gaussian beam, phase singularity

## 1. Introduction

Singular points (or screw phase dislocations) of propagating optical-vortex (OV) [1] beams describe the 3D trajectories that are determined by the amplitude-phase distribution of the field [2]. At these points the amplitude is equal to zero, and the phase is indeterminate [3]. Such trajectories, being an inherent part of the entire optical field, constitute separated smooth structures called the singular skeleton [4]. The singular skeleton can include not only the migration trajectories of singular points (OV trajectory), but also any features and parameters of the field near these points such as vorticity, ellipticity and inclination (morphology parameters) of the OV's cores [4–6]. The singularities are not completely independent; each singularity is coherent with adjacent singularities and organizes the field nearby. Therefore, being the feature of an optical field, the singular skeleton can provide a general notion of the beam spatial profile and its properties (for instance, as in [7,8]). That is why the inspection of the field singularities can contribute to the solution of the "inverse problem of optics": recovering the full amplitude and phase spatial profile from a limited set of the observed singular points [9,10]. Such an approach involving the singular skeleton is undeniably convenient since it is only necessary to know the positions and properties of phase dislocations in individual sections of the optical field.

Edge diffraction of optical vortices has been extensively discussed [11,12,21,22,13–20]. Basically, the study of edge diffraction is relevant for metrological measurements due to the fact that the OV acts as unique and stable marks in the optical field [23,24]. One of the most remarkable features of the edge diffraction of beams with OV can be noted as the phase singularity restoration after a beam is screened [12]. When the OVs are not stopped and are observed in a certain diffracted beam cross section (observation plane), they move along intricate spiral-like trajectories while the screen edge moves across the incident beam. The spiral trajectories for LG and Kummer beams of topological charges 1, 2, and 3 were studied in details [17,18]; the similar 3D spiral trajectories can be observed within the propagating diffracted OV beam in space when the screen position is fixed [19,20]. In [17–20], singular skeletons and their properties were considered, the conditions for the appearance of topological reactions (the rapid movement of singular points – the OV "jumps"), the appearance of loops and pulsations of the OV migration trajectories were discovered and clarified. In this paper, the knowledge about this phenomenon is supplemented and the research interest is focused on identifying the dependences of the optical field parameters on the OV trajectory in the case of the movable screen edge. Such trajectories were also investigated in more extravagant cases: for pulsed broadband THz radiation [25], when the distribution of singular points is simultaneously projected onto the observation plane. In that rate, the study of such beams is of a great interest, in terms of great opportunities for spectrally-resolved holographic registration of amplitude-phase distributions [26–31].

However, in the fundamental research on the subject under discussion [17–20], our attention was paid to the nature disclosure of OV circular beams edge diffraction, and their behavior. Moreover, we revealed the existence conditions of topological reactions, wherein exact dependence on OV beam parameters was ignored. A mathematically clearly defined concept of the weak diffraction perturbation (WDP) was not previously introduced. It was interpreted as "…*Practically, this implies that the beam visibly retains its initial circular shape immediately after the screen, which for low-order LG beams occurs if the screen is displaced from the axis by two or more beam radii measured at $e^{-1}$ intensity level…*"[20]. Such interpretation requires to be supplemented for a more rigorous understanding of the applicability limits between WDP and strong diffraction perturbation (SDP). The OV of the diffracted beam was also established to perform the spiral motion with the screen-edge translation across the incident beam cross section (the process of screening). However, there was no clear dependence of the spiral trajectory configuration on the diffracting singular beam parameters. The listed above circumstances lead to the problem statement of finding universal function describing the OV migrations trajectories. Such function provides the possibility of restoring the field parameters from the known singular skeleton, for example, by determining the function correlation with the OV coordinates of the investigated singular skeleton. In this work, we establish the describing trajectory function for the simplest task: the edge diffraction of a single-charged Laguerre-Gaussian (LG) beam with zero radial index outside the waist, when the OV describes a spiral trajectory on the stationary observation plane and the screen edge performs a translatory motion over the beam cross section from the transverse periphery to the central axis.

We believe that the use of such a basic model object as a half-plane partially screened the beam in a single-beam scheme together with the results obtained in this paper can simplify the problems in some metrological applications. Such problems are currently being solved using double-beam interferometers that are sensitive to vibrations. Moreover, the research results can be useful for such practical tasks as optical telecommunications [32,33]. The advantages of the approach are complemented by the fact that, in the case of opaque apertures, the diffraction affects predictable on the beam amplitude, regardless of its phase component. When there is no possibility of synthesizing coherent reference waves, single-beam diffraction technique can replace the interference approaches, what is especially relevant for analyzing broadband or polychromatic beams. [34]. Although the monochromatic radiation in the visible range is modeled in this paper, the results can be generalized for other frequency ranges with the restriction for Fresnel diffraction. Thus, the expansion of fundamental knowledge about diffraction lays the foundation for the new solutions in relevant problems.

The paper is organized as follows: Section 2 describes amplitude-phase distribution of the diffracted LG beam. Parametric function of spiral trajectories is introduced in Section 3. Section 4 derives the function of spiral trajectories. Section 5 reveals the features of spiral trajectories.

## 2. Amplitude-phase distribution of the diffracted LG beam

Circular symmetric beams with optical vortices can be differ by the forms of the radial field distributions determined by the methods of their synthesis [35]. Due to the creation simplicity and the application breadth, the circular vortex beams described by the LG modes attract attention first. For the deep understanding of the OV edge diffraction, the need is felt for the full disclosure of nature in the case of the simplest beam with a unit topological charge. In this work, we consider the LG beam with positive unit azimuthal and zero radial indexes, the complex amplitude [3] can be represented, with omitted constant amplitude factor, as:

$$E_{LG}(x,y,z_0) = E_0 \cdot \frac{w_0}{w(z_0)} \cdot \frac{(x+iy)\sqrt{2}}{w(z_0)} \cdot \exp\left(-\frac{x^2+y^2}{w(z_0)^2}\right) \cdot \exp\left(i\left(kz + \frac{k(x^2+y^2)}{2R(z_0)} - \arctan\left(\frac{y}{x}\right)\right)\right), \quad (1)$$

where $E_0$ is the amplitude parameter; $x$, $y$ are transversal coordinates; $w_0$ is transversal beam dimension in the waist; $w(z_0) = w_0 \cdot \sqrt{1+z_0^2/z_R^2}$ is the transversal beam dimension on the distance $z_0$; $z_R = \pi \cdot w_0^2 / \lambda$ is the Rayleigh length; $\lambda = 633$ nm is a wavelength of modeled light; $k$ is the wave number; $R(z_0) = z_0 \cdot \left(1 + z_R^2/z_0^2\right)$ is the radius of the wavefront curvature; and $z_0$ is the distance between the beam waist and the screen.

The scheme of edge diffraction of the LG beam is shown in Figure 1. The value $a$ is the distance between the central Z axis and the edge of the moveable screen along X axis, this screen is the perpendicular to Y axis. The observation plane (Figure 1) is the plane where the diffraction pattern is recorded. The value $z$ is the distance between the screen and the observation plane.

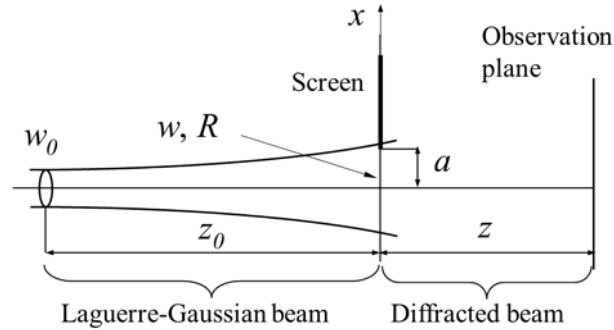

**Figure 1.** Scheme of edge diffraction of the single charged LG beam.

The solution of the diffraction integral for edge diffracted LG beam in the Fresnel approximation has been discussed in various papers [15,36], in our case the integral can be expressed as:

$$E_{DLG}(x,y,z) = \int_{-\infty}^{a}\int_{-\infty}^{+\infty} E_{LG}(x',y',z_0) \cdot \exp\left(\frac{ik\left((x-x')^2+(y-y')^2\right)}{2z}\right) dx'dy', \quad (2)$$

where $x'$, $y'$ are transverse LG beam coordinates in the plane of the screen.

In this work, we use the same solution as in [15,36], but it is written in other notation and up to the real amplitude coefficient:

$$E_{DLG}(x,y,z) = \exp\left(-\left(\frac{q}{2\sqrt{p}}x + a\sqrt{p}\right)^2\right) + \frac{q}{2\sqrt{p}}\sqrt{\pi}\cdot(x+iy)\left(\text{erf}\left(\frac{q}{2\sqrt{p}}x + a\sqrt{p}\right)+1\right), \quad (3)$$

where parameters have the following form $p(z,z_0) = \left( w(z_0)^{-2} - \dfrac{ik}{2} \cdot \left( R(z_0)^{-1} + z^{-1} \right) \right)$, $q(z) = \dfrac{ik}{z}$.

The solution of the diffraction integral (3) gives the intensity and phase distributions of two different types (Figure 2). The intensity can resemble both interference fringes with a fork structure (Figure 2(a)) and a curved smooth donut shape (Figure 2(c)). In addition, either a "wavy" structure (Figure 2(b)) or a smooth irregular structure (Figure 2(d)) also appears in the distribution of the phase. This effect becomes clear if we consider the diffraction pattern as a superposition of an unperturbed incident singular beam and an edge wave which imaginary source is the screen edge [18–20].

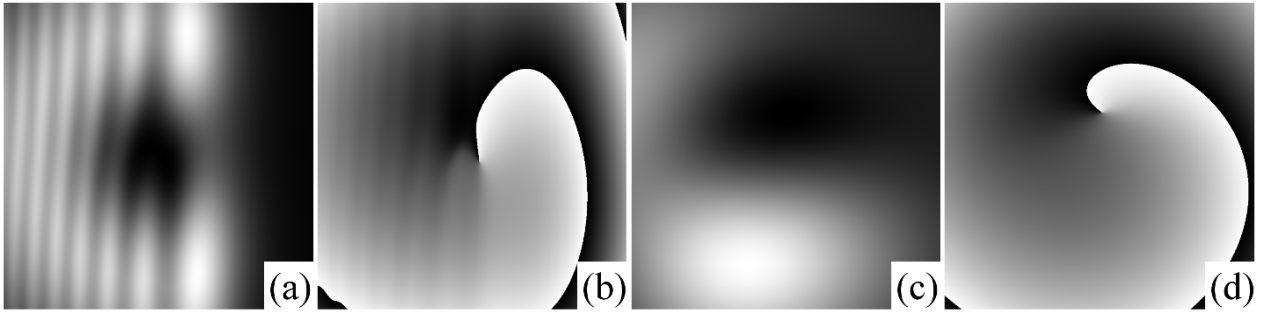

**Figure 2.** Distribution of the intensity (a,c) and the phase (b,d) for edge diffracted LG beam with parameters $z_0=3z_R$, $z=0.5z_R$ (a,b) and $z_0=0.5z_R$, $z=3z_R$ (c,d). The edge position $a$ in both cases is $1.25w$.

Possessing the amplitude-phase distribution of the field (3) like data matrices, it becomes possible to numerically determine the location of the OV points using our program. The algorithm is aimed at searching in the distribution matrix the intensity minima where phase circulation occurs. The accuracy of determining the location is limited to the size of one matrix pixel. By successively screening (decreasing the value $a$), the set of singular points is obtained that form the OV migration trajectories. These sets are below called as numerical trajectories. They serve as control checks in the derivation and verification of equations. This numerical modeling is noted to have demonstrated good agreement with physical experiments especially for single-charged beams [5,17].

**3. Parametric function of spiral trajectories**
Firstly, our aim is to determine the coordinates of the OV in accordance with the position of the screen edge. The OV position coincides with the point where expression (2) is equal to zero. However, the presence of the error function (erf) makes the analysis inappropriate. Therefore, we replace the erf function with hyperbolic tangent (tanh); in this approximation (equation (4)), the good convergence remains for about $a>0.1w$.

$$\exp\left(-\left(\frac{q}{2\sqrt{p}}x+a\sqrt{p}\right)^2\right) + \frac{q}{2\sqrt{p}}\sqrt{\pi}(x+iy)\left(\tanh\left(\frac{2}{\sqrt{\pi}}\left(\frac{q}{2\sqrt{p}}x+a\sqrt{p}\right)\right)+1\right) = 0. \quad (4)$$

The equation (4) is converted to:

$$\exp\left(-\left(\frac{q}{2\sqrt{p}}x+a\sqrt{p}\right)^2\right) \cdot \left[1+\exp\left(\frac{-4}{\sqrt{\pi}}\left(\frac{q}{2\sqrt{p}}x+a\sqrt{p}\right)\right)\right] + \frac{q}{2\sqrt{p}}\sqrt{\pi} \cdot (x+iy) = 0. \quad (5)$$

Now, to separate the real and imaginary parts, we express the equation (5) with new parameters:

$$\exp\left(-\left((r+i\cdot m)x+(r_0+i\cdot m_0)a\right)^2\right) \cdot \left[1+\exp\left(\frac{-4}{\sqrt{\pi}}\left((r+i\cdot m)x+(r_0+i\cdot m_0)a\right)\right)\right] + (r+i\cdot m)\sqrt{\pi} \cdot (x+iy) = 0, \quad (6)$$

where

$$r_0 = \text{Re}\left(\sqrt{p}\right) = \sqrt{\frac{\sqrt{\frac{1}{w^4} + \frac{k^2}{4R_z^2}} + \frac{1}{w^2}}{2}}, \quad m_0 = \text{Im}\left(\sqrt{p}\right) = -\sqrt{\frac{\sqrt{\frac{1}{w^4} + \frac{k^2}{4R_z^2}} - \frac{1}{w^2}}{2}},$$

$$R_z = \left(R(z_0)^{-1} + z^{-1}\right)^{-1},$$

$$r = \text{Re}\left(\frac{q}{2\sqrt{p}}\right) = \frac{m_0 \cdot k/(2z)}{r_0^2 + m_0^2}, \quad m = \text{Im}\left(\frac{q}{2\sqrt{p}}\right) = \frac{r_0 \cdot k/(2z)}{r_0^2 + m_0^2}.$$

Thus, the characteristics of the spiral trajectory are determined by two pairs of introduced parameters ($r$, $m$, $r_0$, $m_0$), which ratio is presented in equation (7). The numerical calculations confirm that the spiral shape remains unchanged for various diffraction conditions at a constant value of $s$.

$$s = -\frac{m}{r} = -\frac{r_0}{m_0}, \tag{7}$$

The diffraction observation distance, the wave number, the transverse parameter, and the radius of the wavefront curvature of the screened beam reveal parameter $s$:

$$s = \sqrt{\sigma^{-2} + 1} + \sigma^{-1}, \tag{8}$$

where

$$\sigma = \frac{kw^2}{2}\left(\frac{1}{R(z_0)} + \frac{1}{z}\right). \tag{9}$$

It should be noted that the equation (8) has the reversal expression: $\sigma = 2\left(s - \frac{1}{s}\right)^{-1}$.

The shape of the spiral is completely determined by the degree of the spiral twist $\sigma$, and its scale can vary depending on the established conditions of the edge diffracted beam. Although parameter $s$ is a derivative of the degree of twist $\sigma$, both parameters will be used to reduce the notation. In addition, these parameters have different physical meanings, which are interpreted below. The behavior of the spiral trajectories was analyzed for $\sigma$ in region [0.917, 23.22] (for $s$ in region [1.044, 2.571]; this parameter is specified in parallel).

After applying the Euler formula to equation (6), it is clearly separated into two ones. The possibility of isolating the real and imaginary parts was the only intention in approximating the error function. The system of equations is obtained in the form:

$$\begin{cases} \tilde{A}_1\left(\cos\varphi_1 + \tilde{A}_2\cos(\varphi_1 + \varphi_2)\right) + 2\sqrt{\pi}(rx - my) = 0 \\ \tilde{A}_1\left(\sin\varphi_1 + \tilde{A}_2\sin(\varphi_1 + \varphi_2)\right) + 2\sqrt{\pi}(rx + my) = 0 \end{cases}, \tag{10}$$

where $\tilde{A}_1 = \exp\left[-\left((r^2 - m^2)x^2 + (r_0^2 - m_0^2)a^2\right)\right]$, $\tilde{A}_2 = \exp\left[-\frac{4}{\sqrt{\pi}}(rx + r_0 a)\right]$, $\varphi_1 = -2(rx + r_0 a)(mx + m_0 a)$,

$\varphi_2 = -\frac{4}{\sqrt{\pi}}(mx + m_0 a)$.

Next, eliminate the variable $y$ from equations system (10):

$$\tilde{A}_1\left(\sin\left(\varphi_1 + \text{atan}\left(\frac{r}{m}\right)\right) + \tilde{A}_2 \sin\left(\varphi_1 + \varphi_2 + \text{atan}\left(\frac{r}{m}\right)\right)\right) + 2x\sqrt{\pi}\sqrt{r^2 + m^2} = 0. \tag{11}$$

The equation (9) is written in quadratic approximation as:

$$Ax^2 + Bx + C = 0, \tag{12}$$

where

$A = \tilde{A}_1 \cdot \left[-2rm \cdot \cos f_1 - 2\left(\frac{k}{2z}a\right)^2\right] + \tilde{A}_2 \cdot \left[\left(-2rm - \frac{4r}{\sqrt{\pi}}\left(\frac{-k}{z}a + \frac{-4m}{\sqrt{\pi}}\right)\right) \cdot \cos f_2 - \frac{1}{2}\left(\frac{-k}{z}a + \frac{-4m}{\sqrt{\pi}}\right)^2 \sin f_2\right],$

$B = 2\sqrt{\pi}\sqrt{r^2 + m^2} + \tilde{A}_1 \cdot \left[\frac{-k}{z}a \cdot \cos f_1 + \tilde{A}_2 \cdot \left[\left(\frac{-k}{z}a + \frac{-4m}{\sqrt{\pi}}\right) \cdot \cos f_2 + \frac{-4r}{\sqrt{\pi}}\sin f_2\right]\right],$

$C = \tilde{A}_1 \cdot \left[\sin f_1 + \tilde{A}_2 \cdot \sin f_2\right],$

and
$$f_1 = \tilde{\varphi}_1 + \operatorname{atan}\left(\frac{r}{m}\right), \quad f_2 = \tilde{\varphi}_1 + \tilde{\varphi}_2 + \operatorname{atan}\left(\frac{r}{m}\right),$$

$$\tilde{A}_1 = \exp\left[-\left(r_0^2 - m_0^2\right)a^2\right], \quad \tilde{A}_2 = \exp\left[-\frac{4}{\sqrt{\pi}}(r_0 a)\right],$$

$$\tilde{\varphi}_1 = -2r_0 m_0 a^2, \quad \tilde{\varphi}_2 = -\frac{4}{\sqrt{\pi}}(m_0 a).$$

The dependence of the coordinate *x* on *a* can be described by formulas with different degrees of approximation (accuracy decreases in the order of formula number):

$$x_0 = \frac{-B + \sqrt{B^2 - 4AC}}{2A}, \tag{13.1}$$

$$x_0 = \frac{BC}{AC - B^2}, \tag{13.2}$$

$$x_0 = -\frac{B}{C}, \tag{13.3}$$

$$x_0 = \frac{C}{\sqrt{4\pi(r^2 + m^2)}}. \tag{13.4}$$

The most accurate solution can be achieved by extracting *x* and *y* from equations (10) (the last term) like in equation (11) and substituting expression (13.1) in the part containing exponential and trigonometric functions, i.e. by substituting the quadratic approximation (13.1) in the initial functions (10). Then final parametric description of the OV position in Cartesian coordinates can be expressed through the system of equations (14). In order to reduce the size of the expression, the same operation can be performed for less accurate approximations (13.2-13.4), they give a higher degree of deviation from the true value, especially in the SDP zone. The fact that, when we use approximations (13.2-13.4), it does not significantly change the general form of the solution. Moreover, it allows drawing some intermediate conclusions about the composition of the function types which is necessary to describe the vortex migration: exponents and cosines (sinuses).

$$\begin{cases} x(x_0, a) = \dfrac{A_1}{\sqrt{4\pi(r^2 + m^2)}}\left(\sin\varphi_1 + A_2 \cdot \sin\left(\varphi_1 + \varphi_2 + \operatorname{atan}\left(\dfrac{r}{m}\right)\right)\right) \\ y(x_0, a) = \dfrac{A_1}{\sqrt{4\pi(r^2 + m^2)}}\left(\cos\varphi_1 + A_2 \cdot \cos\left(\varphi_1 + \varphi_2 - \operatorname{atan}\left(\dfrac{m}{r}\right)\right)\right) \end{cases}. \tag{14}$$

Comparative graphs of numerical calculations and analytical formula (14) describing the position of the vortex on the observation plane depending on the screen position *a* are shown in Figure 3, where the values are normalized by the transverse parameter of the beam in the screen region to represent the beam screened degree more clearly. In the case of a negative azimuthal index of the LG beam, it is necessary to invert the sign in the second equation of system (14).

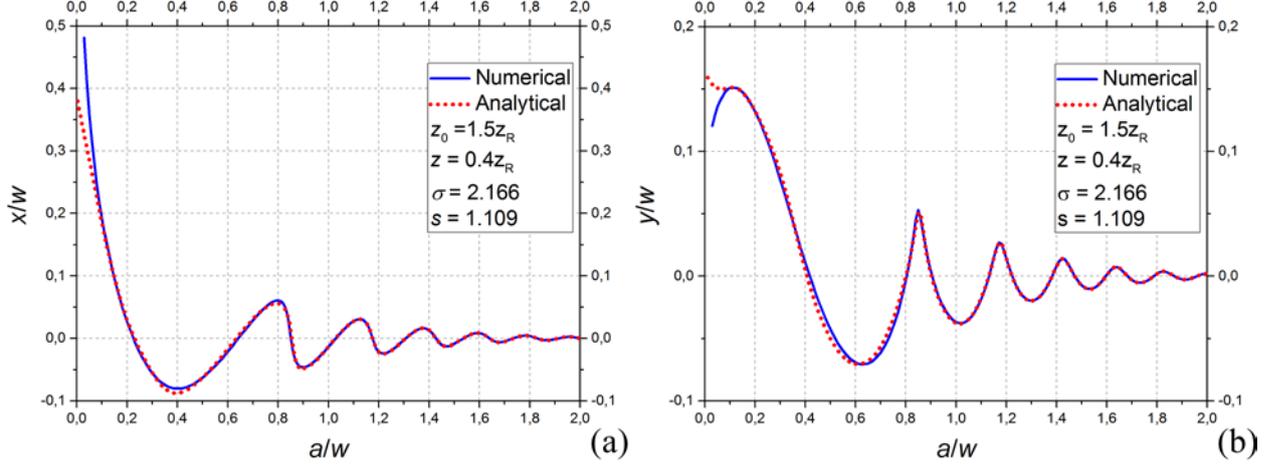

**Figure 3.** Numerical (solid blue line) and analytical (dotted red line) calculations of the coordinate OV position on the screen edge a for the single charged LG beam (parameters are indicated on the graph legend). The boundary between the zones of the WDP and the SDP is approximately at $a = 0.6w$.

## 4. Function of spiral trajectories

In this Section we come to the description of the spiral trajectory that the OV outlines when the LG beam is sequentially screened from its edge to the central axis. It is very convenient to record such a trajectory experimentally since there is no need to monitor the beam screened position *a*.

The parametric dependences of the OV position in polar coordinates (15) and (16) can be derived from system (14).

$$\rho(x,a) = \frac{A_1(x,a)}{\sqrt{4\pi(r^2 + m^2)}} \sqrt{1 + A_2(x,a)^2 + 2 \cdot A_2(x,a)\cos(\varphi_2(x,a))}, \quad (15)$$

$$\varphi(x,a) = \frac{\pi}{2} + \varphi_1 + \operatorname{atan}\left(\frac{r}{m}\right) + \operatorname{atan}\left(\frac{A_2 \sin \varphi_2}{1 + A_2 \sin \varphi_2}\right). \quad (16)$$

Note that the first multiplier in equation (15) prevails in the WDP zone, and the expression under the square root in this zone is about unity.

The direct derivation from equations (15) and (16) of $\rho(\varphi)$ (the OV trajectory described by sequential beam screening) is not a trivial task. It seems possible to find an approximate solution when drawing up the connection of structural elements $(rx+r_0 a)$ and $(mx+m_0 a)$. For this, we use the following assumption, which is especially satisfied well for the WDP:

$$mx + m_0 a \approx \omega \frac{m_0}{r_0}, \quad (17)$$

where $\omega = (rx + r_0 a)$ is the structural element of equation (10).

The last terms ($A_2$ multiplied with sinus and cosines) in equations (14) are extremely small in the WDP zone and can be equated to zero. Then the relationship between the polar angle and the structural element from the equation (10) is determined by the following formula:

$$\varphi \approx -2\omega^2 \frac{m_0}{r_0} + \left(\frac{\pi}{2} + \operatorname{atan}\left(\frac{r}{m}\right)\right), \quad (18)$$

then

$$\omega = \sqrt{\frac{\varphi - \left(\frac{\pi}{2} + \operatorname{atan}\left(\frac{r}{m}\right)\right)}{2} \cdot \frac{-r_0}{m_0}}. \quad (19).$$

The line along which the trajectory oscillates (20) can be obtained by equations (15), (17), (19) and written as:

$$\rho_{WDP}(\varphi) = \frac{1}{\sqrt{\pi}} \cdot \frac{2z}{kw} \sqrt{\sigma^2 + 1} \exp\left[-\frac{\varphi - \mathrm{atan}(s)}{\sigma}\right]. \tag{20}$$

Based on the structure of the equations, we can assume that the complete spiral equation has the multiplication form of the found equation (20) and some unknown functions $f(\varphi)$:

$$\rho(\varphi) = \rho_{WDP}(\varphi) \cdot (1 + f(\varphi)). \tag{21}$$

Next, we will select the unknown function $f(\varphi)$ for the best matching to describe the complete equation of the spiral in the above range of values $s$. If we neglect the presence of oscillations, then the spiral equation can be approximately expressed by two exponential functions:

$$\rho_{\exp}(\varphi) = \rho_{WDP}(\varphi)\left[1 + \frac{1}{s}\exp\left[-\frac{\varphi - \mathrm{atan}(s)}{s\sqrt{\pi}}\right]\right]. \tag{22}$$

To consider the oscillations appearing in the numeric simulation, additional functions containing cosines are added and some adjustment is made to equation (22). Then, the selected with best matching function for eight values $\sigma$ can be represented as:

$$\rho_{osc}(\varphi) = \rho_{WDP}(\varphi) \cdot \left[1 + \frac{\pi + s}{e}\exp\left(-\frac{\varphi - \mathrm{atan}(s)}{s\sqrt{\pi}}\right) - \frac{\pi}{4}\left(s - \frac{3}{4}\right)^2 \exp\left(-\sqrt{2}(s-1)\frac{\varphi - \mathrm{atan}(s)}{2}\right)\cos(\varphi) + \right.$$
$$\left. + \left(0.8311\left(s - \sqrt{\frac{2}{\sqrt{\pi}}}\right)^2 + 0.2376\right) \cdot \exp\left(-\frac{\pi}{2}\left(s - \frac{8}{3\pi}\right)\frac{\varphi - \mathrm{atan}(s)}{2}\right)\cos\left(\varphi - \frac{\pi}{16}\right)\right] \tag{23}$$

In equation (23), the first term with cosine describes the oscillations in the entire surveyed zone, and the second begins to influence closer to the SDP zone. Understanding the true nature of the spiral trajectory is complicated by the presence of the oscillation composition (or possibly oscillations of one complex function). The exponentially damped harmonic oscillations can be only observed in the WDP far zone (with a decrease in the degree of screening). In the case of a negative azimuthal index of the LG beam, it is necessary to replace $\varphi$ with $-\varphi$ in equations (20-23) and start a count of $\varphi$ in the negative region of real values ($\varphi<0$).

Figure 4 shows the graph of the trajectory deviations in polar coordinates for the spiral twist $\sigma = 0.917, 2.166, 9.648$ and $23.22$ ($s = 2.571, 1.563, 1.109$ and $1.044$ responsible).

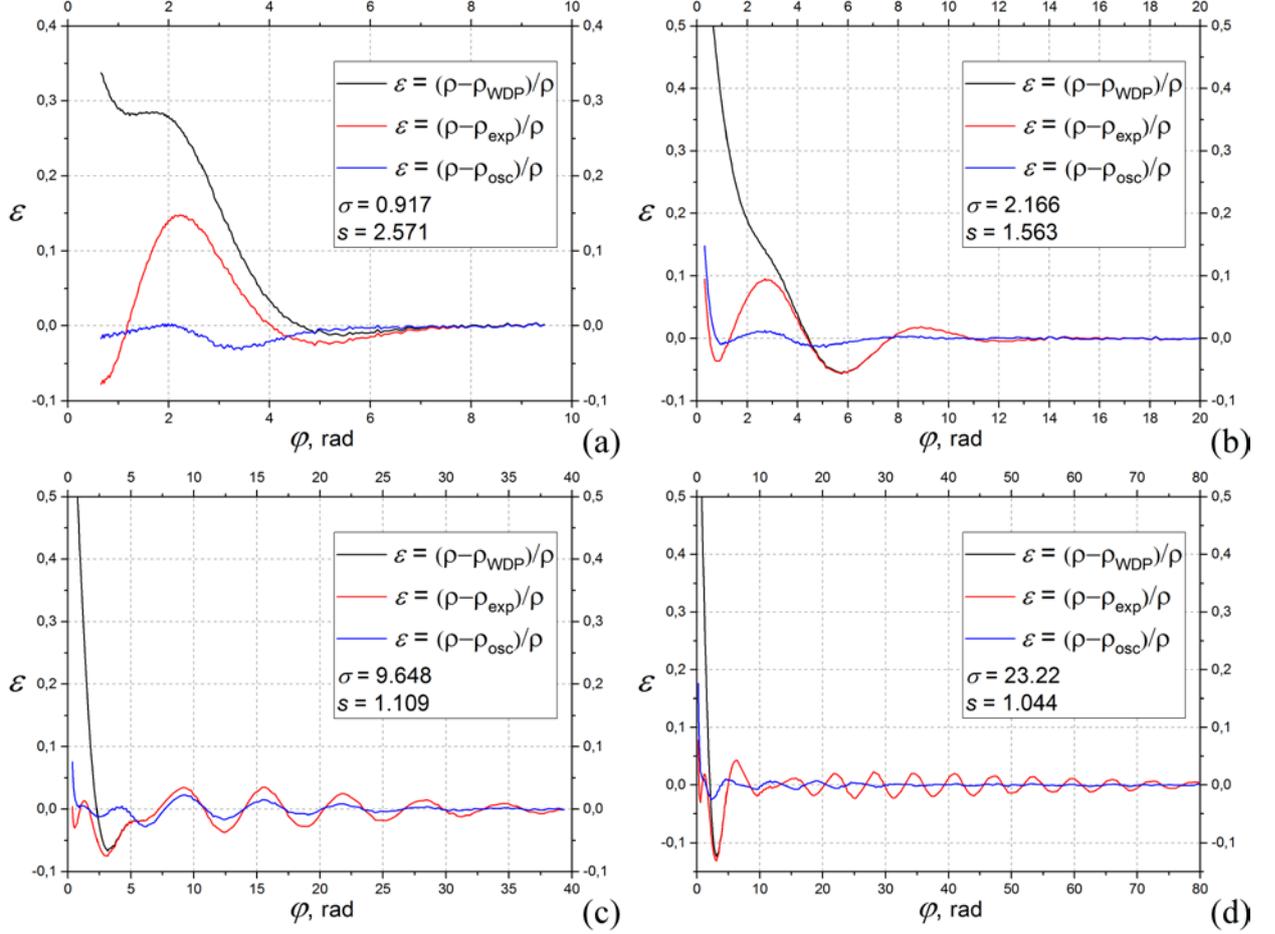

**Figure 4.** Deviation of analytical calculations (equations (20), (22), (23)) relative to numerical. Diffraction parameters are displayed on graphs. The black line is partially hidden under the red line. Insignificant fluctuations are caused by an error in numerical modeling of $\rho$.

The equation (21) of the spiral trajectory in polar coordinates is controlled by many parameters and is especially sensitive for small values of the angle $\varphi$ (the greatest distance of the singular point from the central axis) due to the influence of exponents in the third and fourth terms. Nevertheless, the selected function retains quite satisfactory accuracy in the chosen interval of the spiral twist parameter $s$. Namely, the deviations from numerical calculations are up to 2-4% for equation (21) and up to 6-15% for equation (20) in range $\sigma$ of [0.917, 23.22] and an angle $\varphi$ greater than 0.6-0.9 radians. At small $\varphi$ (the screen position $a$ is close to zero), the selected functions demonstrate no adequate result in terms of converging with numerical simulation. The selected function (21) becomes less accurate for the values of $s$ outside the researching range. For instance, when $\sigma=0.352$ ($s=5.845$), the deviation is up to 80%, and for formula (20) it increases only up to 22%.

It should be also noted that when the parameter $s$ is changed, a general rotation of the spiral is observed, which is apparently controlled by the magnitude atan($s$). Although not as expressive as $\sigma$ (due to a slight affect in the angle), the derived parameter $s$ can also be considered as characteristic value.

## 5. Features of spiral trajectories

In this paper the OV migration for the fixed observation planes along the propagation axis with the screen transitional overlaps the LG beam is investigated; the distance from the central axis to the screen edge $a$ decreases from 2-4$w$ to 0, for large values of $a$ the diffracted beam remains practically unperturbed. In the screening process, the OV makes a spiral motion, which form depends on the curvature radius of the wave front $R$, the transverse parameter of the beam $w(z_0)$, and the observation distance $z$. Depending on the ratio of these values, the twist of the spiral changes so that we can clearly observe both several coils (Figure 5(a)) and less than one (Figure 5(b)). It is noteworthy that for the single charged LG beam the "jumps" of OV are also revealed as for multicharged vortex beams [18–20]. This phenomenon happens as follows: the OV dipole is born at a remote point, and then the initial singular point collides with the oppositely charged point of this dipole and annihilates with it, after which only one OV remains. Such a process proceeds extremely quickly and can be perceived as the "jump" of only one OV. For the case shown in Figure 5(a), the "jump" criterion is fulfilled [19]. In each coil of the spiral with a "jump" the value of $M$ ($M = \frac{k\rho a}{z}$) remains greater than unity ($M$=1.07, 1.18, 1.15 marked on graph). At the points of discontinuity of the spiral trajectory (which marked with stars in Figure 5(a)), there are regions of the OV dipole birth and annihilation.

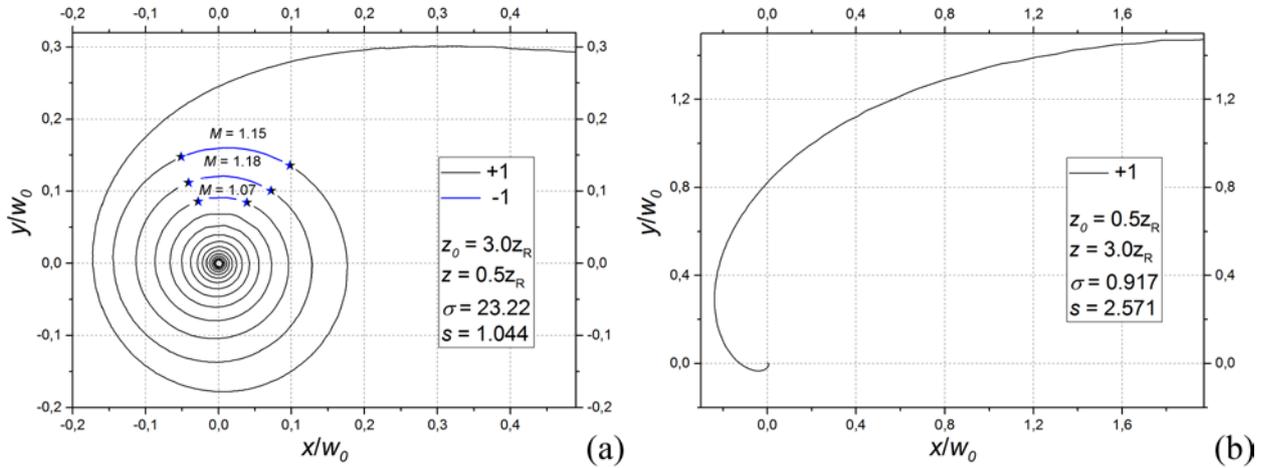

**Figure 5.** Spiral-like OV trajectories for the edge diffracted single charged LG beam for $\sigma$=23.22 (a) and $\sigma$=0.917 (b) (the black line for positive and the blue line for negative charged OV). Diffraction beam parameters are displayed on graphs.

Spiral trajectories are difficult to be analyzed and they give no understanding which particular function can be expressed. The representation of these trajectories in polar coordinates $\rho(\varphi)$ resembles an extremely sharp hyperbolic function, which is also inappropriate approach due to the loss of accuracy in representation and the decrease in sensitivity to changes in diffraction parameters. Anyway, such a sharp function can be countered by another one; if the representation in polar coordinates uses the logarithmic scale for the radial distance, then the trajectories become close to the straight lines (Figure 6). Some features immediately appear in this system, namely, the presence of a linear part for WDP [18], which transforms into bend under SDP – the rest of the non-linear part, and the presence of small oscillations along the entire trajectory (oscillations are present for all values $\sigma$, which determines their amplitude).

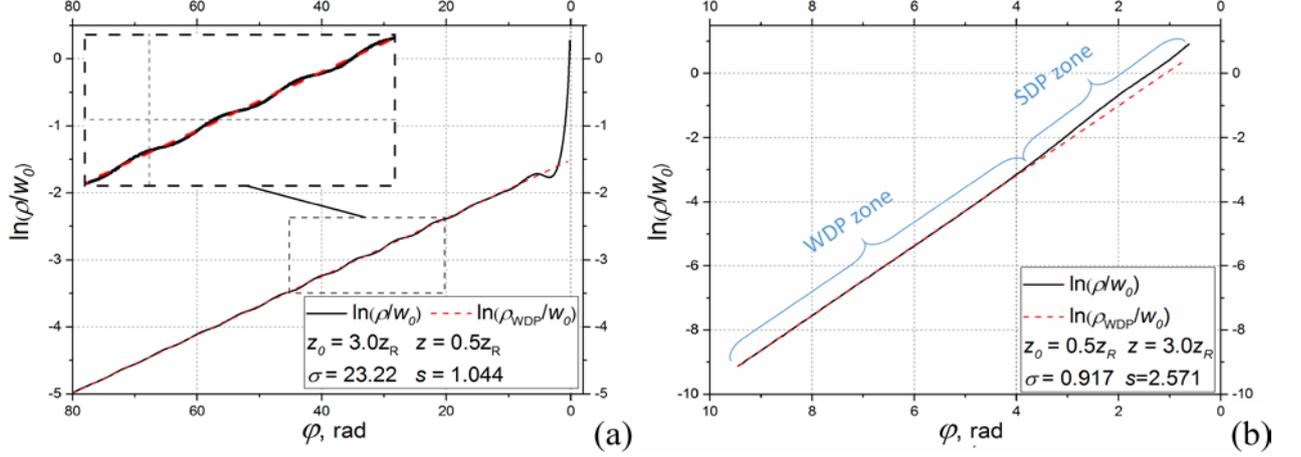

**Figure 6.** OV trajectories in the logarithmic polar coordinate system for $\sigma=23.22$ (a) and $\sigma=0.917$ (b). Diffraction beam parameters are displayed on graphs.

Since spirals have more than one coil, an understanding of the polar angle reference should be introduced. We assume that with the increasing degree of the screening, the angle $\varphi$ decreases from an infinitely large value to zero, when the OV is maximally removed from the initial position. Therefore, the horizontal coordinate axis in Figure 6 is directed downward $\varphi$ from a certain initial value. Representation of the OV trajectory in polar coordinates with a logarithmic scale gives reason to perform some classification of the OV edge diffraction phenomenon: for the WDP the trajectory oscillates slightly along the straight line (dashed line in Figure 6, it was mathematically described by the equation (20)), and for the SDP it has a pronounced deviation from this line. Thus, the WDP can be represented as an exponential dependence of the polar radii from the polar angle. The distinction between these two zones is quite arbitrary, since being determined by the degree of assumption at which the deviation from the line remains acceptable. Nevertheless, there is some tendency to narrow the SDP zone with decreasing parameter of the spiral twist $\sigma$. The boundary between the zones is located on the last coil of the spiral; the corresponding transition angle is about 4 radians; a more accurate location can range from $\pi$ to $3\pi/2$. If we consider the location of the screen edge at which segregation occurs, the position of the edge varies significantly from $0.34w$ for $\sigma=23.22$ to $1.7w$ for $\sigma=0.917$.

The parameter $\sigma$ plays a decisive role in the form of the resulting spiral trajectory. This means that the number of registered coils can indicate its value. Knowing that the SDP affects only the last half-turn of the spiral, the approximate number of coils is expressed from the equation for WDP (20) as:

$$n = \sigma \frac{\ln(\rho_1/\rho_2)}{2\pi}, \qquad (24)$$

where $\rho_1$ and $\rho_2$ are the radial distances between which the number of coils $n$ is calculated.

From equation (24), we see a direct proportional relationship between $n$ and $\sigma$. Theoretically, the spiral contains an infinite number of coils, which requires us to introduce the following assumption of the trajectory visibility limits. If $\rho_1$ is defined as the distance $\rho$ for the angle $\pi/2$ (i.e. $\rho_1 = \rho(\pi/2)$), and $\rho_2 = \rho_1/\exp(\pi) = \rho_1/23.1$, then $\sigma$ determines the number of visible half-turns of the spiral trajectory.

Having carefully considered the equation (9), such a parallel is drawn. In the case of the infinitely large radius of the wavefront curvature of the screened beam, the parameter $\sigma$ becomes equivalent to the Fresnel number. This demonstrates the global relationship of the OV behavior with the optical system parameters and can provide a new approach to determining the beam parameters.

## 6. Conclusions

The research presented above contains plenty of details that reveal the edge diffraction phenomenon of the single charged Laguerre-Gaussian beam. One of these is the illustration of two different reaction rates of the OV to the displacement of the straight edge relative to the beam axis. These two reaction rates can be conditionally separated by two zones of diffraction perturbation: WDP and SDP. The first zone is characterized by an exponential dependence of the polar coordinates of the OV position with the insignificant oscillations presence, which decreases as the OV comes to the central axis and has a period of the polar angle equal to $2\pi$ (for all types of spirals). The second zone (SDP) is characterized by a higher sensitivity to the displacement of the screen edge and is indescribable by one exponential function as for the WDP (logarithm does not reveal the linear dependence).

One of the discovered features of spiral trajectories is that they can be self-similar to each other up to a scaling factor for various beam parameters. The type of spiral or its degree of twist is determined by the parameter $\sigma$, which in turn depends on the radius of wave front curvature, the beam transverse size, and the observation distance of the diffraction pattern. The founded parameter $\sigma$ can be interpreted as the number of visible half-turns in the spiral trajectory. That can be useful in a preliminary (fast) search for the parameters of the diffracting LG beam. In addition, the parameter $\sigma$ can be likened to the Fresnel number. This fact shows a deep physical connection between the behavior of the singular phase dislocation and the general propagation of the optical field.

In our work the functions of the OV position in 3D space (equation (14)) are derived. They establish the dependence of the transverse coordinates on the observation distance and the screening degree. Based on this, the 2D trajectory equations (20) and (22-23) are obtained, which gives a possibility to describe the OV position in the cross section during the sequential edge beam screening. The edge is displaced from 2-4w to the beam center, so that the entire area of the OV migration, where changes occurred, covered.

The assumptions have been made during the study provided the analytical analysis, which, although not to the full extent, led to the derivation of equations describing the OV migration. Consequently, the equation (14) is derived that establishes the relationship between the OV location (in Cartesian coordinates) and the position of the screen edge. The exact equation of the line where the OV migrates along, with a shifting screen in the WDP zone, is established. Based on the analytical calculations, the spiral trajectory function is selected with best matching. Its applicability remains limited for the coefficient of spiral twist $\sigma = 0.917 \ldots 23.22$ and requires a more thorough study. However, this selected function shows the general nature of the trajectory and can become a support for further research.

This paper is the first of its kind, where the development of the analytical approach to the determination of the trajectories has begun. The presented parametric equations (14) provide the opportunity to describe the migration of the singularity in 3D space for various degrees of the beam blocking by fixed screen. The universality of these equations is determined by all diffraction parameters, including the observation distance and the radiation wavelength. They are applicable for both WDP and SDP zones with the exception for very strong beam blocking (approximately $a < 0.1w$). In addition, this approach can be extended to the case of edge diffracted broadband beams, which individual temporal spectrum components form the spectral spiral trajectory in a cross section. Thus, the half blocked in the focal waist broadband vortex beam and the formation of OV trajectory slightly resembling a spiral is examined in [25]. Such spectral trajectories have a similar nature as the studied ones. However, based on the results of the present work, we can assume that if we consider the edge screened broadband vortex beam in the WDP zone, then the spiral forms with many coils, which provides more information about beam diffraction. Finally, the knowledge of the OV trajectory (and skeleton), that takes place in the case of partial blocking of the vortex beam, allows us to accurately localize the obstacle. Such a problem statement is relevant in ensuring the noise immunity of wireless telecommunication systems [32,33]. To summarize, the results of our research contribute to solving the inverse diffraction problem – determining the parameters of the optical field from the OV trajectory.

**Acknowledgements**

The work presented here was funded by Russian Science Foundation, Project No 19-72-10147.

**References**

[1] Shen Y, Wang X, Xie Z, Min C, Fu X, Liu Q, et al. Optical vortices 30 years on: OAM manipulation from topological charge to multiple singularities. Light Sci Appl 2019;8:90. https://doi.org/10.1038/s41377-019-0194-2.

[2] Leach J, Dennis MR, Courtial J, Padgett MJ. Vortex knots in light. New J Phys 2005;7:55–55. https://doi.org/10.1088/1367-2630/7/1/055.

[3] Soskin MS, Vasnetsov MV. Singular optics. Prog. Opt., 2001, p. 219–76. https://doi.org/10.1016/S0079-6638(01)80018-4.

[4] Dennis MR, O'Holleran K, Padgett MJ. Chapter 5 Singular Optics: Optical Vortices and Polarization Singularities. Prog. Opt., 2009, p. 293–363. https://doi.org/10.1016/S0079-6638(08)00205-9.

[5] Khoroshun A, Chernykh O, Tatarchenko H, Sato S, Kozawa Y, Popiołek-Masajada A, et al. Chain of optical vortices synthesized by a Gaussian beam and the double-phase-ramp converter. OSA Contin 2019;2:320. https://doi.org/10.1364/OSAC.2.000320.

[6] Izdebskaya Y. Optical necklaces generated by the diffraction on a stack of dielectric wedges. Phys Lett A 2008;372:3909–13. https://doi.org/10.1016/j.physleta.2008.02.068.

[7] Szatkowski M, Popiołek Masajada A, Masajada J. Optical vortex trajectory as a merit function for spatial light modulator correction. Opt Lasers Eng 2019;118:1–6. https://doi.org/10.1016/j.optlaseng.2019.01.014.

[8] Popiołek-Masajada A, Sokolenko B, Augustyniak I, Masajada J, Khoroshun A, Bacia M. Optical Vortex Scanning in an aperture limited system. Opt Lasers Eng 2014;55:105–12. https://doi.org/10.1016/j.optlaseng.2013.10.023.

[9] Volostnikov VG. Phase problem in optics. J Sov Laser Res 1990;11:601–26. https://doi.org/10.1007/BF01120784.

[10] Baranova N, Zel'dovich B. Dislocations of the wave-front surface and zeros of the amplitude. Zh Eksp Teor Fiz 1981;80:1789–97.

[11] Marienko IG, Soskin MS, Vasnetsov M V. Diffraction of optical vortices. In: Angelsky O V., editor. Fourth Int. Conf. Correl. Opt., 1999, p. 27. https://doi.org/10.1117/12.370421.

[12] Vasnetsov M V., Marienko IG, Soskin MS. Self-reconstruction of an optical vortex. J Exp Theor Phys Lett 2000;71:130–3. https://doi.org/10.1134/1.568297.

[13] Gorshkov VN, Kononenko AN, Soskin MS. Diffraction and self-restoration of a severely screened vortex beam. In: Soskin MS, Vasnetsov M V., editors. Second Int. Conf. Singul. Opt. (Optical Vor. Fundam. Appl., 2001, p. 127–37. https://doi.org/10.1117/12.428259.

[14] Bekshaev A, Mohammed KA. Transverse energy redistribution upon edge diffraction of a paraxial laser beam with optical vortex. In: Angelsky O V., editor. Elev. Int. Conf. Correl. Opt., 2013, p. 906602. https://doi.org/10.1117/12.2050875.

[15] Bekshaev AY, Mohammed KA, Kurka IA. Transverse energy circulation and the edge diffraction of an optical vortex beam. Appl Opt 2014;53:B27. https://doi.org/10.1364/AO.53.000B27.

[16] Bekshaev AY, Mohammed KA. Spatial profile and singularities of the edge-diffracted beam with a multicharged optical vortex. Opt Commun 2015;341:284–94. https://doi.org/10.1016/j.optcom.2014.12.019.

[17] Chernykh A, Bekshaev A, Khoroshun A, Mikhaylovskaya L, Akhmerov A, Mohammed KA. Edge diffraction of optical-vortex beams formed by means of the fork hologram. In: Angelsky O V., editor. Twelfth Int. Conf. Correl. Opt., vol. 9809, 2015, p. 980902.

https://doi.org/10.1117/12.2220398.

[18] Bekshaev A, Chernykh A, Khoroshun A, Mikhaylovskaya L. Localization and migration of phase singularities in the edge-diffracted optical-vortex beams. J Opt 2016;18:024011. https://doi.org/10.1088/2040-8978/18/2/024011.

[19] Bekshaev A, Chernykh A, Khoroshun A, Mikhaylovskaya L. Singular skeleton evolution and topological reactions in edge-diffracted circular optical-vortex beams. Opt Commun 2017;397:72–83. https://doi.org/10.1016/j.optcom.2017.03.062.

[20] Bekshaev A, Chernykh A, Khoroshun A, Mikhaylovskaya L. Displacements and evolution of optical vortices in edge-diffracted Laguerre–Gaussian beams. J Opt 2017;19:055605. https://doi.org/10.1088/2040-8986/aa6352.

[21] Cvijetic N, Milione G, Ip E, Wang T. Detecting Lateral Motion using Light's Orbital Angular Momentum. Sci Rep 2015;5:15422. https://doi.org/10.1038/srep15422.

[22] Giovanni Milione GM, Ting Wang TW, Jing Han JH, and Lianfa Bai and LB. Remotely sensing an object's rotational orientation using the orbital angular momentum of light (Invited Paper). Chinese Opt Lett 2017;15:030012–6. https://doi.org/10.3788/COL201715.030012.

[23] Wang W, Yokozeki T, Ishijima R, Wada A, Miyamoto Y, Takeda M, et al. Optical vortex metrology for nanometric speckle displacement measurement. Opt Express 2006;14:120. https://doi.org/10.1364/OPEX.14.000120.

[24] Wang W, Yokozeki T, Ishijima R, Takeda M, Hanson SG. Optical vortex metrology based on the core structures of phase singularities in Laguerre-Gauss transform of a speckle pattern. Opt Express 2006;14:10195. https://doi.org/10.1364/OE.14.010195.

[25] Kulya M, Semenova V, Gorodetsky A, Bespalov VG, Petrov N V. Spatio-temporal and spatiospectral metrology of terahertz broadband uniformly topologically charged vortex beams. Appl Opt 2019;58:A90. https://doi.org/10.1364/AO.58.000A90.

[26] Petrov N V., Kulya MS, Tsypkin AN, Bespalov VG, Gorodetsky A. Application of Terahertz Pulse Time-Domain Holography for Phase Imaging. IEEE Trans Terahertz Sci Technol 2016;6:464–72. https://doi.org/10.1109/TTHZ.2016.2530938.

[27] Kulya MS, Semenova VA, Bespalov VG, Petrov N V. On terahertz pulsed broadband Gauss-Bessel beam free-space propagation. Sci Rep 2018;8:1390. https://doi.org/10.1038/s41598-018-19830-z.

[28] Balbekin NS, Kulya MS, Belashov A V., Gorodetsky A, Petrov N V. Increasing the resolution of the reconstructed image in terahertz pulse time-domain holography. Sci Rep 2019;9:180. https://doi.org/10.1038/s41598-018-36642-3.

[29] Kulya M, Petrov N V., Tsypkin A, Egiazarian K, Katkovnik V. Hyperspectral data denoising for terahertz pulse time-domain holography. Opt Express 2019. https://doi.org/10.1364/oe.27.018456.

[30] Kulya M, Petrov N V., Katkovnik V, Egiazarian K. Terahertz pulse time-domain holography with balance detection: complex-domain sparse imaging. Appl Opt 2019;58:G61. https://doi.org/10.1364/AO.58.000G61.

[31] Kulya MS, Balbekin NS, Gorodetsky AA, Kozlov SA, Petrov N V. Vectorial terahertz pulse time-domain holography for broadband optical wavefront sensing. In: Sadwick LP, Yang T, editors. Terahertz, RF, Millimeter, Submillimeter-Wave Technol. Appl. XIII, SPIE; 2020, p. 12. https://doi.org/10.1117/12.2547714.

[32] Mendoza-Hernández J, Arroyo-Carrasco ML, Iturbe-Castillo MD, Chávez-Cerda S. Laguerre–Gauss beams versus Bessel beams showdown: peer comparison. Opt Lett 2015;40:3739. https://doi.org/10.1364/OL.40.003739.

[33] Ahmed N, Zhao Z, Li L, Huang H, Lavery MPJ, Liao P, et al. Mode-Division-Multiplexing of Multiple Bessel-Gaussian Beams Carrying Orbital-Angular-Momentum for Obstruction-Tolerant Free-Space Optical and Millimetre-Wave Communication Links. Sci Rep

[34] Soskin MS, Polyanskii P V., Arkhelyuk OO. Computer-synthesized hologram-based rainbow optical vortices. New J Phys 2004;6:196–196. https://doi.org/10.1088/1367-2630/6/1/196.
[35] Wang X, Nie Z, Liang Y, Wang J, Li T, Jia B. Recent advances on optical vortex generation. Nanophotonics 2018;7:1533–56. https://doi.org/10.1515/nanoph-2018-0072.
[36] Masajada J. Half-plane diffraction in the case of Gaussian beams containing an optical vortex. Opt Commun 2000;175:289–94. https://doi.org/10.1016/S0030-4018(00)00470-3.

2016;6:22082. https://doi.org/10.1038/srep22082.